# Effect of charge modulation in $(LaVO_3)_m(SrVO_3)_n$ superlattices on the insulator-metal transition

W. C. Sheets, B. Mercey, and W. Prellier[*]

*Laboratoire CRISMAT, CNRS UMR 6508, ENSICAEN,*
*6 Blvd Maréchal Juin, F-14050 Caen Cedex, France*

(Dated: 12 October 2007)

**Abstract**

A series of epitaxial $(LaVO_3)_{6m}(SrVO_3)_m$ superlattices having the same nominal composition as $La_{6/7}Sr_{1/7}VO_3$, a Mott-Hubbard insulator, were grown with pulsed-laser deposition on [001]-oriented $SrTiO_3$ substrates, and their superlattice period was varied. When $m = 1$, the insulating resistivity of bulk-like $La_{6/7}Sr_{1/7}VO_3$ is obtained; however, an increase in the periodicity ($m \geq 2$) results in metallic samples. Comparison of the superlattice periodicity with the coherence length of charge carriers in perovskite oxide heterostructures are used to understand these observations. A filling-controlled insulator-metal transition was induced by placing a single dopant layer of $SrVO_3$ within $LaVO_3$ layers of varying thickness.

PACS: 81.15.Fg; 71.30.+h; 73.61.Ng

---

[*] wilfrid.prellier@ensicaen.fr

The properties of perovskite oxides containing multiple A or B-site cations of different oxidation state are, to a great extent, determined by the cation and charge ordering. Thin film methods offer a versatile approach to overcome the natural preference for disorder or low-dimensional ordering in certain materials by controlling the location of cations in tailored oxide multilayers.[1] Fabrication of such heterostructures, composed of different perovskite oxides, has been utilized to explore the change in, for example, the magnetic response, magneto-resistance, and ferroelectric properties.[2-4] Artificial heterostructures have been utilized more recently to probe how interlayer coupling and charge ordering at the interface affect such properties.[5-7] For instance, Ohtomo and coworkers deposited a single layer of $LaTiO_3$ in a $SrTiO_3$ matrix.[8] This $LaTiO_3/SrTiO_3$ superlattice exhibits metallic conductivity, similar to a bulk solid solution of $La_{1-x}Sr_xTiO_3$, even though the heterostructure is based on two insulators. The spatial distribution of $Ti^{3+}$ across the atomically abrupt $LaTiO_3/SrTiO_3$ interfaces was mapped by electron energy-loss spectroscopy (EELS), revealing that a crossover in Ti oxidation states (from 3+ to 4+) occurs over a length of 10 Å. Another study on the interface of a $LaVO_3/LaVO_4$ superlattice revealed the presence of a two-dimensional layer of $V^{4+}$ between the two layers, corresponding to a $LaVO_x$ phase with an oxidation state that does not exist in the bulk.[9] It is anticipated, therefore, that an interface between layers of $LaVO_3$ and $SrVO_3$ in a heterostructure will yield a mixed $V^{3+}$ and $V^{4+}$ valence. It is of particular interest to know the length scale of this interaction and its effect on the electrical transport properties of $LaVO_3/SrVO_3$ superlattices, including whether varying the periodicity of such superlattices can induce a filling controlled transition from a Mott-Hubbard insulating to a metallic state. In this contribution, successful syntheses of varying period superlattices of epitaxial $(LaVO_3)_m(SrVO3)_n$ thin films, in which the electrical transport properties are affected noticeably by the superlattice period $\Lambda$, are presented.

Rare-earth vanadates containing $V^{3+}$, such as $LaVO_3$ and $YVO_3$, present a challenge to band theories of electron behavior in the solid state. Since their 3$d$-levels are partially filled ($d^2$ configuration), straightforward application of the molecular orbital band theory to these materials predicts electron de-localization in partially filled bands and, therefore, metallic properties. Instead, these materials show insulating, not metallic behavior, implying that the $d$-electrons are localized in discrete, atomic-like states. Such materials are called Mott-Hubbard insulators. In particular, the $(La,Sr)VO_3$ system has been studied extensively owing to its classic filling-controlled insulator-metal transition behavior.[10,11] The solid solution $La_{1-x}Sr_xVO_3$ exists over the entire composition ($0 \leq x \leq 1$) all the way from $LaVO_3$, a Mott-Hubbard insulator that undergoes an antiferromagnetic (AFM) transition, to $SrVO_3$, a paramagnetic metallic conductor, without disrupting the distorted perovskite structural network. On introducing $Sr^{2+}$ into $LaVO_3$, a mixed $V^{3+}$ and $V^{4+}$ valence occurs, which favors the formation of $\sigma^*$ and $\pi^*$ bands whose broadening eventually results in metallic behavior.[12] Indeed, the Mott-Hubbard insulating phase disappears at a strontium substitution level of $x > 0.2$, and doping with more holes produces a metallic state.[11,13-15] Although $La_{1-x}Sr_xVO_3$ samples are metallic for $x > 0.2$, the AFM phase persists until $x > 0.26$, after which samples become paramagnetic metals.

The $(LaVO_3)_m(SrVO_3)_n$ superlattice films were deposited on [001]-oriented $SrTiO_3$ using PLD from single phase, ceramic targets of $LaVO_4$ and $Sr_2V_2O_7$. The cubic lattice parameter of the $SrTiO_3$ substrate $a_{STO}=3.905$ Å matches well with the pseudocubic cubic lattice parameter of $LaVO_3$ (LVO), $a_{LVO}=3.92$ Å[16] and cubic lattice parameter of $SrVO_3$ (SVO), $a_{SVO}=3.843$ Å.[17] Superlattices of $(LaVO_3)_m(SrVO_3)_n$ were grown at a substrate temperature of 700 °C and oxygen pressure of $10^{-5}$ Torr, conditions which represented the best compromise for simultaneously stabilizing both SVO and LVO phases in a heterostructure. Several important deposition parameters were: KrF excimer laser with $\lambda = 248$ nm, pulse rate = 3 Hz, laser power density $\approx 2$ J cm$^{-2}$, and target to substrate distance = 75 mm. Deposition rates of 0.2 and 0.4 Å/pulse for LVO and SVO, respectively, were calibrated under these conditions. Post-growth the samples were cooled to room-temperature under dynamic vacuum ($10^{-5}$ Torr) at a rate of 10 °C min$^{-1}$. The crystalline structure of the thin film samples was examined by X-ray diffraction (XRD) using a Seifert 3000P diffractometer (Cu K$_\alpha$, $\lambda = 1.5406$ Å).

To probe the length scale of charge screening at the interface between LVO and SVO layers a series of $(LaVO_3)_{6m}(SrVO_3)_m$ thin films samples, where the overall stoichiometry is equal to $La_{0.86}Sr_{0.14}VO_3$, were deposited under identical conditions. Each superlattice consisted of depositing $6m$ integer layers of LVO, followed by $m$ integer layers of SVO, repeated in appropriate sequence, such that the total perovskite unit cells grown is an integer near 350. In addition, a $La_{6/7}Sr_{1/7}VO_3$ sample with a superlattice periodicity $\Lambda < a_p$ (where $\Lambda$ is the superlattice period and $a_p$ is the perovskite lattice parameter) also was prepared as a reference solid solution sample. Conventional $\Theta-2\Theta$ scans of the superlattices revealed no extra peaks other than the (00$l$) Bragg reflections of the substrate, constituent film, and satellites peaks. The high-quality of the artificially induced cation order is evidenced by the numerous observed satellite peaks (inset Fig. 1) for each sample. The denoted number $i$ indicates the $i$-th order satellite peak. The average out-of-plane lattice parameter of the superlattice ($a_p$) was obtained from the central (001) Bragg peak of the film and the superlattice period ($\Lambda$) for each sample was determined from the observed $i$-th order peaks about the (001) and (002) film peaks. Figure 1 illustrates that the measured $(LVO)_{6m}(SVO)_m$ superlattice periods $\Lambda$ were in good agreement (< 3 % error) with targeted values. The error in superlattice stacking was quantified by $\Delta = [(\Lambda - (7m \times a_p))/ \Lambda] \times 100$. The lattice parameter $a_p$ was consistent for each superlattice of different periodicity, ranging from 3.945 to 3.952 Å; their similarity to each other and that of the solid solution $La_{6/7}Sr_{1/7}VO_3$ argues for constant stoichiometry of the superlattices.

The overall conductivity of the $(LaVO_3)_{6m}(SrVO_3)_m$ multilayers are a function of the spacing between, and thickness of, the dopant $SrVO_3$ layers. The resistivity of these samples transitions systematically from that of a solid solution to bulk-like individual layers with an increase in periodicity. Four-point resistance measurements were measured by a Quantum Design physical property measurement system (PPMS) and the resistivity was calculated from these measurements using the appropriate correction factors for a two-dimensional rectangular sample with finite thickness.[18] Figure 2a shows the temperature-dependent resistivity of the $(LaVO_3)_{6m}(SrVO_3)_m$ superlattices along with thin film samples of LVO, SVO, and solid solution $La_{6/7}Sr_{1/7}VO_3$ ($\Lambda < a_p$) deposited by PLD under the same conditions. No kinks, which signal the occurrence of a first-order structural phase transition from orthorhombic to monoclinic form,[15,19] are visible in the resistivity plots of the $(LaVO_3)_{6m}(SrVO_3)_m$ superlattice samples. Suppression of this structural phase transformation for thin film samples of the solid solution $La_{1-x}Sr_xVO_3$ has been reported previously.[12] For $La_{1-x}Sr_xVO_3$ samples where the strontium concentration ($x$) is equal to 1/7 or 0.14, the expectation from the electronic phase diagram is that these samples should be AFM and Mott-Hubbard insulators at low temperatures.[15] As shown in Figure 2a, only the $m = 1$ multilayer displays insulating, not metallic, behavior, implying that there is significant interlayer coupling and dispersion of the extra charge carrier within the smallest period heterostructure. Indeed, the magnitude and shape of the resistivity plot for the $m = 1$ superlattice is similar to that measured for the $\Lambda < a_p$ thin film sample, which can be considered to have a statistical distribution of the A-site cations. As mentioned previously, EELS measurements have shown that the charge distribution of an extra charge carrier associated with the single dopant layer naturally spreads beyond the heterostructure interface into the host layers in order to minimize the electronic free energy.[8] At a single $LaTiO_3/SrTiO_3$ interface, the length scale of the charge interaction was measured to be 10 Å for the crossover between valence states, while the total spatial distribution of the dopant charge carriers was ~40 Å. Such distances are consistent with the $m = 1$ multilayer ($\Lambda = 27.2$ Å) exhibiting a resistivity similar to that of a solid solution. In contrast, no insulator-metal transition is observed for superlattice samples with $m > 1$ and, indeed, their resistivity are significantly lower. For these superlattices, their periodicity (54.7 - 188.9 Å) exceeds the coherence length of the charge carriers, thereby localizing electron holes within the SVO layer and near the LVO/SVO interface. Hence, the hole concentration exceeds the critical dopant value ($x > 0.2$) within these layers and the samples are metallic. As shown in Figure 2b, the resistivity of the $m > 1$ superlattices remains metallic until there is an upturn at low temperatures. Similar low temperature resistivity upturns have been reported for La$_{1-x}$Sr$_x$VO$_3$ samples in the critical region of the insulator-metal transition, $0.178 < x < 0.26$ for bulk crystals[15] and $0.20 < x < 0.28$ for thin films.[12] As the superlattice periodicity increases, however, there is a systematic decrease in the onset temperature of the resistivity upturn (Fig. 2b), indicating a gradual transition to the metallic SVO phase. In a LaTiO$_3$/SrTiO$_3$ multilayer, a minimum thickness of five LaTiO$_3$ layers was required for the center titanium site to recover bulk-like spectroscopic features.[8] Nonetheless, a complete transition in the resistivity of the samples to that of bulk SVO was not expected because the presence of insulating layers between metallic SVO layers increases the resistivity of a multilayer sample.[6]

As seen in figure 3, deposition of a single layer of SVO embedded within LVO layers of varying thickness results in a filling controlled insulator-metal transition. Each (LaVO$_3$)$_m$(SrVO$_3$) superlattice consisted of depositing $m$ integer layers of LVO where $1 \leq m \leq 7$, followed by a single layer of SVO ($n = 1$), repeated in appropriate sequence, such that the total perovskite unit cells grown is an integer near 350. The (LaVO$_3$)$_m$(SrVO$_3$) superlattices where $m \geq 5$ ($x \leq 0.18$) are Mott-Hubbard insulators. At high temperature (Fig. 4a), the conduction mechanism of these samples obeys Arrhenius activated behavior with activation energies ($E_a$) of 0.017 ($m = 5$), 0.018 ($m = 6$), and 0.021 eV ($m = 7$) eV. Such values are in close agreement with activation energies reported previously for bulk, 0.020 eV ($x = 0.2$),[13] and thin film, 0.022 eV ($x = 0.15$) and 0.018 eV ($x = 0.175$), La$_{1-x}$Sr$_x$VO$_3$ samples.[12] When $m \leq 4$ ($x \geq 0.2$), a filling-controlled transition across the insulator-metal phase boundary occurs and the samples become metallic, which is in accord with the results of previous studies on bulk solid solutions of La$_{1-x}$Sr$_x$VO$_3$.[14] As illustrated in figure 4b, apart from the upturns at low temperature for $m = 3$ and 4 samples, the temperature (T) dependence of resistivity ($\rho$) for the metallic superlattice samples ($m = 1, 2, 3,$ and 4) can be well expressed up to 375 K by the relation $\rho \approx \rho_o + A \cdot T^2$, where $\rho_o$ represents the residual resistivity and $A$ the transport coefficient in a Fermi liquid. The coefficients $A$ were calculated to be on the order of $10^{-9}$ Ω·cm K$^{-2}$ (Fig. 4b), consistent with a strongly correlated electron system, and are also enhanced as the Mott-Hubbard insulator is approached from the metallic side, 2.97 ($m = 1$), 3.25 ($m = 2$), 4.41 ($m = 3$), and 5.18 n$\Omega$·cm K$^{-2}$ ($m = 4$).[20] Below the calculated effective Fermi temperatures ($T^* \approx 450$ K), the product $A(T^*)^2/a$, where $a$ is the in-plane lattice parameter of the sample, is less than or remains on the order of $h/e^2$, as expected in a Fermi liquid where the transport is dominated by electronic correlations. It should be noted that a non-Fermi liquid $T^{1.5}$ dependence for the resistivity of bulk $La_{1-x}Sr_xVO_3$ samples near the insulator-metal phase boundary has been reported previously, which the authors interpreted was due to AFM spin fluctuations at the critical boundary between an AFM and paramagnetic metal.[14,21] When the resistivity of the metallic $La_{1-x}Sr_xVO_3$ superlattices are fit to a $T^{1.5}$ dependence, the calculated coefficients $A$ for the plots are on the order of $10^{-7}$ $\Omega$·cm K$^{-1.5}$ and increase as the Mott-Hubbard insulating phase is approached from the metallic side. Both the magnitude and trend in the coefficients $A$ for a $T^{1.5}$ dependence are consistent with what has been reported previously for bulk samples of $La_{1-x}Sr_xVO_3$.[14,15]

In summary, a series of high-quality $(LaVO_3)_{6m}(SrVO_3)_m$ superlattices with a constant $La_{6/7}Sr_{1/7}VO_3$ stoichiometry have been fabricated using standard pulsed-laser deposition, and the thickness of the subcells was varied. The as-deposited film samples display either Mott-Hubbard insulating bulk-like ($m = 1$) or metallic ($m \geq 2$) resistivity depending on the value of the superlattice period. The insulator-metal transition can be explained by comparing the superlattice periodicity with the known coherence length of charge carriers across the interface of perovskite oxide heterostructures. This interlayer coupling was then used to induce a gradual filling controlled insulator-metal transition when a single layer of SVO was embedded as a dopant within LVO layers of varying thickness. In general, these results demonstrate the utility of PLD to obtain LVO/SVO heterostructures with high crystalline order and to modify their electrical properties by a change in superlattice periodicity, including dopants confined to a single layer.

The authors thank J. Lecourt and L. Gouleuf who aided in the experiments and Dr. R. Frésard for helpful discussions. This work was carried out in the frame of the NoE FAME (FP6-5001159-1), the STREP MaCoMuFi (NMP3-CT-2006-033221), and the STREP CoMePhS (NMP4-CT-2005-517039) supported by the European Community and by the CNRS, France. Partial support from the ANR (NT05-1-45177, NT05-3-41793) is also acknowledged. W.C.S was additionally supported through a Chateaubriand post-doctoral fellowship.

Figure Captions.

Figure 1. A plot illustrates the evolution of the calculated periodicity ($\Lambda$) as a function of the targeted value ($7m \times a_p$) for several $(LaVO_3)_{6m}(SrVO_3)_m$ superlattices. The line is drawn as a guide to the eye. The inset shows a $\Theta-2\Theta$ pattern recorded around the (001) reflection of $SrTiO_3$ for a $(LaVO_3)_{18}(SrVO_3)_3$ superlattice with a calculated periodicity of 81.0 Å.

Figure 2. (color online) A graph of the (a) absolute and (b) normalized temperature dependent resistivity of $(LaVO_3)_{6m}(SrVO_3)_m$ superlattices compared with films of $LaVO_3$, $SrVO_3$, and a solid solution $La_{6/7}Sr_{1/7}VO_3$.

Figure 3. (color online) A graph of the temperature dependent resistivity of $(LaVO_3)_m(SrVO_3)_1$ superlattices ($1 \leq m \leq 7$) compared with the end member films of $LaVO_3$ and $SrVO_3$.

Figure 4. (color online) Plots of the (a) Natural logarithm of the resistivity ($\ln \rho$) as a function of the inverse temperature (1000/T), and (b) temperature dependence of resistivity ($\rho$) for $(LaVO_3)_m(SrVO_3)_1$ superlattice samples as a function of $T^2$. The best-fit lines calculated with data points from the linear portion of the resistivity curves are shown.

**Figure 1.**

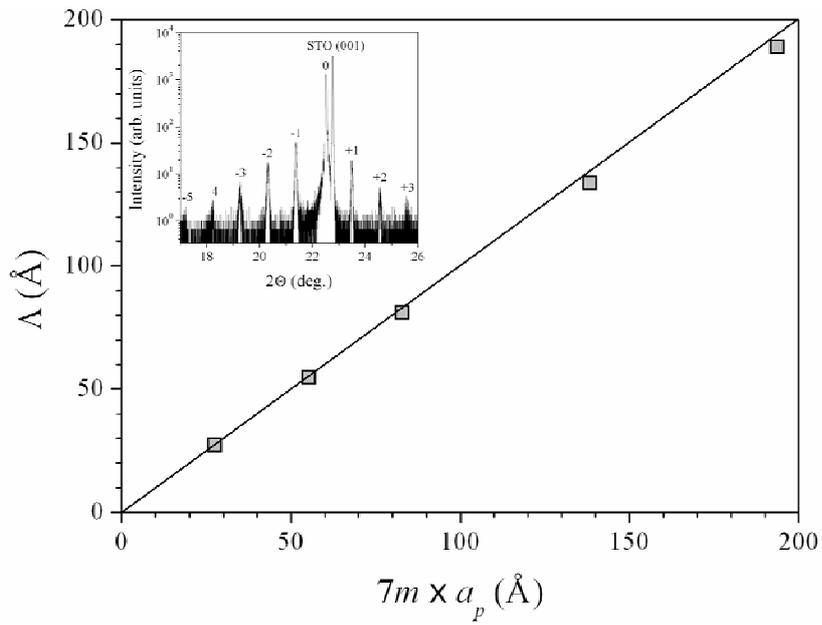

**Figure 2.**

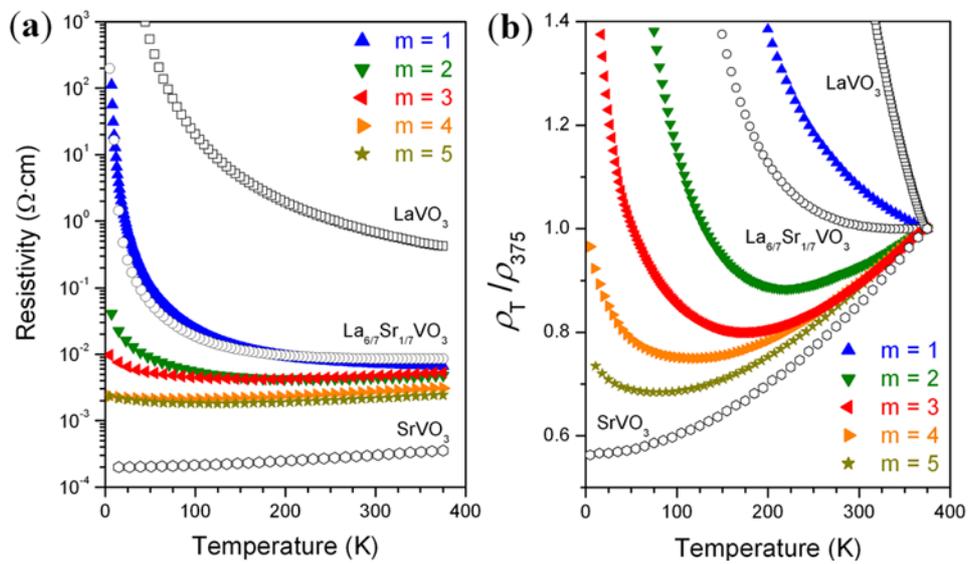

**Figure 3.**

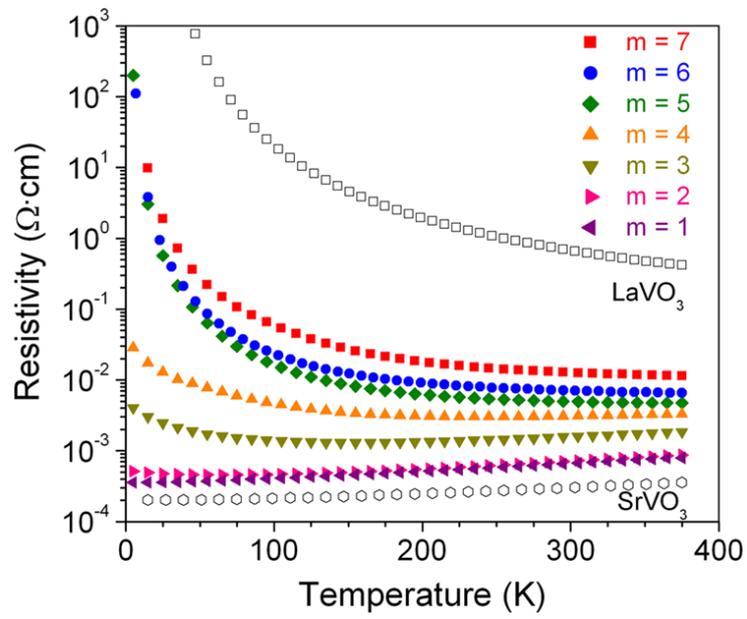

**Figure 4.**

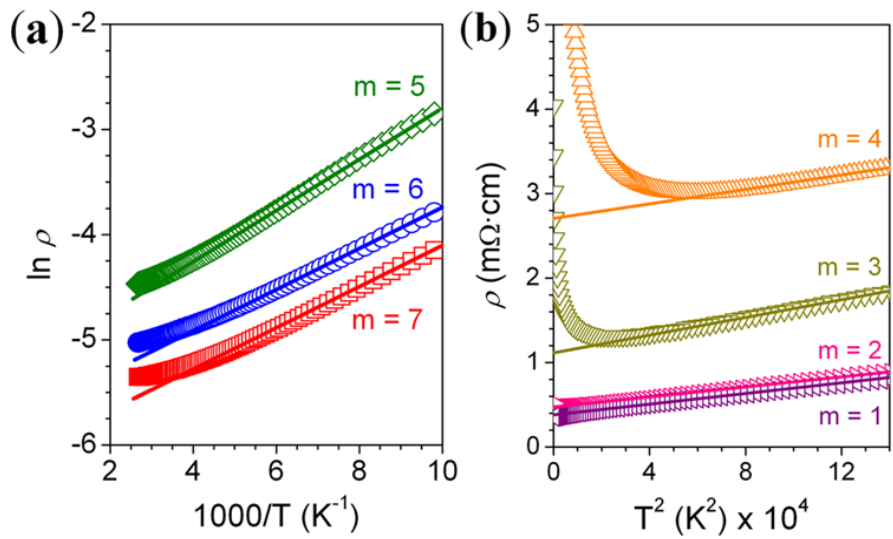